\documentclass[12pt]{article}

\usepackage{graphics,epsfig}
\def\bb#1{\hbox{\mybb#1}}

\def\dalemb#1#2{{\vbox{\hrule height .#2pt
        \hbox{\vrule width.#2pt height#1pt \kern#1pt
                \vrule width.#2pt}
        \hrule height.#2pt}}}

\let\a=\alpha \let\b=\beta \let\g=\gamma \let\d=\delta \let\e=\epsilon
\let\z=\zeta  \let\th=\theta  \let\k=\kappa
\let\l=\lambda \let\m=\mu  \let\x=\xi \let\p=\pi 
\let\s=\sigma \let\t=\tau    
\let\vp=\varphi \let\vep=\varepsilon
\let\w=\omega       \let\D=\Delta \let\Th=\Theta \let\L=\Lambda
 \let\P=\Pi \let\S=\Sigma  
 \let\W=\Omega
\let\la=\label \let\ci=\cite 
  
\def\nn{\nonumber} \def\bd{\begin{document}} \def\ed{\end{document}}
\def\ds{\documentstyle} \let\fr=\frac \let\bl=\bigl \let\br=\bigr
\let\Br=\Bigr \let\Bl=\Bigl
\let\bm=\bibitem
\let\na=\nabla
\def\tU{{\widetilde U}}
\let\pa=\partial \let\ov=\overline
\def\ie{{\it i.e.\ }}
\newcommand{\be}{\begin{equation}}
\newcommand{\ee}{\end{equation}}
\def\ba{\begin{array}}
\def\ea{\end{array}}
\def\ft#1#2{{\textstyle{{\scriptstyle #1}\over {\scriptstyle #2}}}}
\def\fft#1#2{{#1 \over #2}}
\def\F#1#2{{ F_{#1}^{(#2)} }}
\def\cF#1#2{{ {\cal F}_{#1}^{(#2)} }}

\def\={\, =\, }
\def\+{\, +\, }
\def\-{\, -\, }

\def\R{{\bf R}}
\def\sst#1{{\scriptscriptstyle #1}}
\def\oneone{\rlap 1\mkern4mu{\rm l}}
\def\e7{E_{7(+7)}}
\def\td{\tilde}
\def\wtd{\widetilde}
\def\im{{\rm i}}
\newcommand{\ho}[1]{$\, ^{#1}$}
\newcommand{\hoch}[1]{$\, ^{#1}$}
\newcommand{\bea}{\begin{eqnarray}}
\newcommand{\eea}{\end{eqnarray}}
\newcommand{\ra}{\rightarrow}
\newcommand{\lra}{\longrightarrow}
\newcommand{\Lra}{\Leftrightarrow}
\newcommand{\ap}{\alpha^\prime}
\newcommand{\bp}{\tilde \beta^\prime}
\newcommand{\cB}{{\cal B}}
\newcommand{\cO}{{\cal O}}
\newcommand{\vecx}{\vec{x}}
\newcommand{\vecy}{\vec{y}}
\newcommand{\vecp}{\vec{p}}
\newcommand{\vecq}{\vec{q}}
\newcommand{\tr}{{\rm tr} }
\newcommand{\Tr}{{\rm Tr} }

\newcommand{\cL}{{\cal L}}
\newcommand{\cA}{{\cal A}}
\def\ve{\varepsilon}
\def\vf{\varphi}
\def\F{\Phi}
\def\wg{\wedge}

\def \nn {\nonumber}
\def \rk  {m}
\def \L {{\Lambda}}
\def \ka  { {\kappa }}
\def \S {{ \call S}}

\def\up{\uparrow}
\def\down{\downarrow}

\def \foot {\footnote}
\def \bi{\bibitem}

\def \tr {{\rm tr}}
\def \ha {{1 \over 2}}
\def \td {\tilde}
\def \ci{\cite}
\def \N {{\mathcal N}}
\def \ww {\Omega}
\def \const {{\rm const}}
\def \ss {\sum_{i=1}^3 }
\def \t {\tau}
\def\S{{\mathcal S} }
\def \XX {{\rm X}}

\def \lra {\leftrightarrow}
\def \vom {{\bar \omega}}
\def \E {{\mathcal  E}} \def \J {{\mathcal  J}}
\def \YY {{\rm Y}}

\def \d {\del}
\def \rJ {{J}}
\def \sms {sigma models\ }
\def \sm {sigma model\ }
\def \L {\Lambda}
\def \gl {\ell}
\def \tr {{\rm tr\ }}
\def\z{\zeta}
\def\zi{\zeta_1}
\def\zii{\zeta_2}
\def\K{\mbox{K}}
\def\eE{\mbox{E}}   \def \vt {\vartheta}
\def \vr {\varrho}
\def \wup {w}

\def\dg{\dagger}
\def\a{\alpha}
\def\b{\beta}
\def\e{\varepsilon}
\def\p{\phi}
\def\ap{\alpha^\prime}
\def\I{{\cal I}}

\def\xb{{\bar X}}
\def\Tr{{\rm  Tr}}
\def\tr{{\rm  tr}}

\def \del{\partial}
\def \a {\alpha}
\def \aa {{\a'}}
\def\g{\gamma}
\def\s{\sigma}
\def\z{\zeta}
\def\zi{\zeta_1}
\def\zii{\zeta_2}
\def\ov{\over}

\def\I{{\cal I}}
\def\J{{\mathcal J}}
\def \ok {{1\ov \k}}
\def\LL{{\mathcal L }}
\def \jL {{J}}
\def \om {\omega}
\def \cL {{\mathcal L}} \def \cH {{\mathcal H}}
\def\E{{\mathcal E}}
\def\w{\omega}
\def\b{\beta}
\def\l{\lambda}
\def\eps{\epsilon}
\def\vep{\varepsilon}
\def \De {{\mathcal D}}

\def  \Jt {  {J}_{\rm tot}    }

\def \k {\kappa}
\def\foot{\footnote}
\def \four{{\textstyle {1\ov 4}}}
 \def \third { \textstyle {1\ov 3
}}
\def\det{\hbox{det}}
\def \ci {\cite}

\def \foot {\footnote}
\def \bi{\bibitem}

\def \tr {{\rm tr}}
\def \ha {{1 \over 2}}
\def \tid {\tilde}
\def \vv {{\rm v}}
\def \tl {{\tilde \l}}
\def \XX {{\rm X}}
\def \ta {{\tilde \a}}
\def \fo { {1\ov 4}}
\def \ep {\epsilon}
\def \inti {{\int^{2\pi}_0 {d \sigma \ov 2 \pi}}}

\def \d {\partial}
\def \K {{\rm S}}
\def \el {\ell}
\def \Tr {{\rm Tr}}
\def \P {\Phi}
\def \l  {\lambda}
\def \tl {{\tilde \l}}
\def \bl {{\tilde \l}}
\def \const {{\rm const}}
\def \V {v}

\def \bv {v^*}
\def \vv {{\rm v}}
\def \LL {{\mathcal L}}
\newcommand{\PV}[1]{P_{\!\!_{V_{#1}}}}

\def \S {{\rm S}}
\def \vn {\vec n}
\def \tl {\td \l}
\def \td {\tilde}
\def \Prod {\Pi}
\def \O {{\mathcal O}}
\def \Q {{\rm  Q}}
\def \D {\Delta}
\def \N {{\mathcal N}}
\def\tN{{\tilde N}}

\def \m {\mu}
\def \vs {\vec \s}
\def \ie {i.e.}

\def \cD {{\cal D}}

\def  \le  {\l_{\rm eff}}

\def \rS {{\rm S}}
\def\as{{\a}}
\newcommand{\bra}[1]{\mbox{$\langle #1 |$}}
\newcommand{\ket}[1]{\mbox{$| #1 \rangle$}}

\def\e{\epsilon}

\def \bi{\bibitem}
\def \la {\label}

\def \l {\lambda}
\def\foot{\footnote}
\def \tl  {{\tilde \l}}
\def \sql {{\sqrt \l}}
\def \adss {$AdS_5 \times S^5$\ }
\newcommand{\rf}[1]{(\ref{#1})}
\def \ov {\over}

\def\th{\theta}
\def\Th{\Theta}
\def\vth{\vartheta}
\def\btheta{{\bar\theta}}
\def\ttheta{{{\tilde\theta}}}
\def\bttheta{{{\bar\ttheta}}}
\def\vth{\vartheta}

\def\ra{\rightarrow}
\def\N{{\cal N}}
\def\F{{\cal F}}
\def\uM{\underline{M}}
\def\uN{\underline{N}}
\def\uP{\underline{P}}
\def\cc{\circ}
\def\eqv{\equiv}

\def\ni{\noindent}
\def \ha{{1\ov 2}}
\def \bw {{\rm w}}

\def\r{{\rm r}}
\def \rg{{\rm g}}

\def \J {\mathcal{J}}
\def \del {\partial}

\def\dF{\dot{F}}
\def\dG{\dot{G}}
\def\df{\dot{f}}
\def \E {{\cal E}}

\def \S {{\cal S}}
\def \J {{\cal J}}

\def\ms{\mathcal{S}}
\def\mj{\mathcal{J}}
\def\soj{\fr{\ms}{\mj}}
\def \R {{\bf R}}
\def \om {\omega}
\def \tH {\widetilde H}
\def \bE {\bar E}
\def \x {{\cal X}}

 \def \bb {\bar \beta}
\def \W {{\cal E}}

\def \bi{\bibitem}
\def \la {\label}

\def \mm {{\rm m }} 
\def \l {\lambda}
\def\foot{\footnote}
\def \tl  {{\tilde \l}}
\def \sql {{\sqrt \l}}
\def \sqtl {{\sqrt {\tilde \l}}}

\def \HH {{\rm E}}

\def \adss {$AdS_5 \times S^5$\ }

\def \D {\Delta}
\def \thet {\theta}
 \def \t {\tau}
 \def \p {\phi}
 \def \r {\rho}
 \def \rN {{\rm N}}
 \def\tw{{\tilde w}}
 \def\hJ{{J}}
 \def\hw{{w}}
 \def\hl{{\lambda}}
 \def\hth{{\theta}}
 \def\NN{{\cal N}}
 \def \bv {{ \bar w}}
\def \vn {{\vec n}}

\def \ov {\over}

\def \varpi {{\rm w}}
\def \OO {{\cal O}}

\def \rt {{\rm t}}
\def \no {\nonumber }

 \def \bJ {{\rm J}}

\def \adss {$AdS_5 \times S^5$\ }

\begin{document}
\overfullrule=0pt
\parskip=2pt
\parindent=12pt
\headheight=0in \headsep=0in \topmargin=0in \oddsidemargin=0in

\vspace{ -3cm} \thispagestyle{empty} \vspace{-1cm}
\begin{flushright} Imperial-TP-AT-6-6\\
TCDMATH 06-17
\end{flushright}
\begin{center}
 \vspace{2cm}
{\Large\bf
Logarithmic corrections to higher twist scaling \\
\vspace{0.3cm}
 at strong coupling from AdS/CFT}

 \vspace{.5cm} { S. Frolov$^{a,}$\footnote{frolovs@maths.tcd.ie},
  A. Tirziu$^{b,}$\footnote{tirziu@mps.ohio-state.edu}
 and A.A.
 Tseytlin$^{c,}$\footnote{Also at
 Lebedev  Institute, Moscow.\ \
  tseytlin@imperial.ac.uk
 }}\\
 \vskip 0.3cm

{\em $^{a}$  School of Mathematics, Trinity Colege, Dublin 2, Ireland
\\
$^{b}$Department of Physics, The Ohio State University,
Columbus, OH43210, USA\\
\vskip 0.08cm $^{c}$  Blackett Laboratory, Imperial College,
London SW7 2AZ, U.K. }

\end{center}

 \begin{abstract}
We  compute 1-loop correction $E_1$ to the energy of folded  string in $AdS_5
\times S^5$ (carrying spin $S$ in $AdS_5$ and   momentum  $J$ in $S^5$)
using  a ``long
string'' approximation in which  $S \gg J \gg 1$.  According to the AdS/CFT  the function $E_1$
 should represent first subleading correction to strong coupling
expansion of anomalous dimension of  higher twist $SL(2)$ sector operators
 of the form $\Tr D^S Z^J$.
We show that $E_1$  smoothly interpolates  between
the $\ln S$ regime (previously found in the $J\to 0$ case)
and  the $\lambda/J^{2} \ln^3 (S/J)$  regime  (which is the  leading
correction to the  thermodynamic limit on the
spin chain side). This supports
 the universality of the $\ln S$  scaling.  As in the previous
work, we also find ``non-analytic''  corrections  related to
non-trivial 1-loop phase in the corresponding
  Bethe ansatz S-matrix.

\end{abstract}
\newpage

\def \x {\xi}
\def \N {{\cal N}}

\renewcommand{\theequation}{1.\arabic{equation}}
 \setcounter{equation}{0}

\section{Introduction and Summary }

Logarithmic scaling   of anomalous dimensions
of composite operators  with large Lorentz spin  £$S$  is of major importance  in QCD
and was recently also  at the center of attention in planar $\N=4$ SYM
theory in the context of AdS/CFT.
 Ref. \ci{gkp} made a remarkable  observation that
the logarithmic  behaviour
$\Delta=S+ ( k_1 \l + k_2 \l^2+ ...) \ln S +   O(S^0)$
previously known at weak `t Hooft coupling should  continue  also at
strong coupling: the classical energy  of a folded rotating string
in $AdS_5$ which should be dual to a minimal twist operator scales
at large $\l$ and  large
$S\ov \sqrt \l $ as
 $E_0=S+ { \sql \ov \pi} \ln S +   O(S^0)$.

Ref. \ci{ft1} made a  next step by computing the leading quantum
string 1-loop ($1\ov \sql$)
correction $E_1$  to the string energy,  confirming that,
as  at weak coupling,
all terms growing faster than logarithm of $S$  cancel  out
so that one ends up with
$E_1= - {3 \ln 2 \ov \pi}  \ln S  +  O(S^0)$. This provided strong support
to the conjecture that in  planar $\N=4$  SYM  theory
the coefficient  of $\ln S$
(i.e. the  ``scaling function''  or  ``cusp anomalous dimension'' \ci{cusp1,cusp2,cusp3})
should be a  function of $\l$  that smoothly   interpolates between
the weak  and strong coupling regimes
\be\la{sca} \D=E= S + f(\l) \ln S  + O(S^0)  \ . \ee
Arguments  in favour   of such interpolation
 (based on Pade approximations)
 were further  advanced in  \ci{kot}.

Ref.\ci{ft1}  also generalized  the rotating  folded string solution of \ci{dev,gkp}
to the case when the string center of mass is also  moving along big circle of $S^5$
so that the
string is  carrying in addition to the  Lorentz spin $S$  an   $SO(6)$  spin $J$.
Introducing large non-zero $J$  is important in particular since  it
makes it easier  to identify
the corresponding dual gauge theory  operators as belonging to the $SL(2)$
sector \ci{bsb,b} in gauge theory \ci{bfst}:\  $\tr ( D^S Z^J)+...$
(here $D$ is l.c. covariant  derivative and $Z$ is a complex scalar).
While  $J=2$ corresponds to the minimal-twist case, the opposite case
of  large $J$ and small $S$ is a BMN-type \ci{bmn}  limit.
Since $J$ plays the role of the spin chain  length on the gauge theory side
having both $S$ {\it and}   $J$  large is important in order to be
able to apply the thermodynamic limit approximation and,  more generally,
to be able to use the
asymptotic Bethe ansatz of \ci{bds,bs}
in the first place.

The classical string energy for this solution  turned out to be a complicated
function of two arguments $E_0= \sql \E ( \S, \nu),\   \S\equiv { S\ov \sql},
\nu \equiv  { J\ov \sql}$  (given by a solution of two equations involving
Jacobi elliptic functions) but it simplifies  in various special limits.
First, one may consider either ``short''   ($\sqrt{ 1 + \nu^2} \gg 2 \S$)
or ``long'' string.
In the ``short string'' limit  \ci{ft1,bfst}  with  $J \gg S$
one gets the BMN-type scaling $E_0=  J + S \sqrt{1 + {\l\ov J^2} } + ...$.
In the ``long string'' limit $S \gg \sql, \ S \gg J$ and  one should further distinguish the
two cases \ci{ft1}:\foot{Here ``slow'' and ``fast''
  refers to the string  center of mass motion.}

 (i) ``slow long string''   $J \ll \sql \ln {S \ov \sql} $
 in which case $E_0 \approx  S + { \sql \ov \pi} \ln {S} + { \pi J^2 \ov 2 \sql  \ln S }  + ...$,
 i.e. one recovers  the $\ln S$ scaling  of the $J=0$  case
 even though $J$ may still be large in absolute terms since  $S \gg J  \sim  \sql $;

 (i) ``fast  long string''   $S \gg J \gg \sql \ln {S \ov \sql} $
 in which case one finds a familiar ``fast spinning string'' \ci{ft2}  scaling
 $E_0 \approx  S + J[1 +   { \l \ov J^2 } h_1( { S\ov J})  + { \l^2 \ov J^4 }h_2( { S\ov J}) + ...]$
 with $h_1= { 1 \ov 2 \pi^2} \ln^2 { S\ov J}$, $h_2 \sim \ln^4 { S\ov J}$, etc.
 As was shown in \ci{bfst} (see also \ci{kz}), the leading $h_1$ term here is reproduced as
 the corresponding term in the   1-loop anomalous dimension on the gauge theory side,
 implying its non-renormalizability as one goes  from weak to strong coupling in the
 above large charge limit.

\noindent
More recently, ref. \ci{bgk} made  an interesting observation that the expressions for the string
energy in \ci{ft1,bfst} imply that the $\ln^k$-terms appearing
 in the ``fast  long string'' case
 can be resummed
in a closed form: in the
case when   $J \sim   \sql \ln {S \ov \sql} $  the energy can be written as
 \be  \label{as}
 E_0 = S +  J \sqrt{ 1 + { \l \ov \pi^2 J^2 } \ln^2 { S \ov J}  }  + ... \
 =  S +  J \sqrt{ 1 +  { x^2  }  } \  + ...  \ ,  \ee
which is valid when
 \be\la{pp}
 S \gg J \  \ \ \ \ { \rm and}
 \ \ \ \ \  x  \equiv { \sqrt{\l } \ov \pi  J}  \ln { S \ov J} =
  {\rm fixed\ }
   \ .
 \ee
Remarkably, this   formula  captures both  the  ``slow long string'' limit\foot{This resummation is formally true for small $x$ but
the result can then be extended to large $x$ as well, see sect. 2.
As we show in section 2, a natural quantity to be kept fixed in the ``slow long string'' limit is
$ { \sqrt{\l } \ov \pi  J}  \ln {S\ov \sqrt\l} $.  For $S\gg J\gg \sqrt\l$ the difference between this variable and $x$ is negligible.  To have a smooth limit $J\to 0$ one should, however, replace $\ln {S\ov J}$ by $\ln S$ in eqs.\rf{as}-\rf{ouo}.
Let us mention  that a  discussion of the
logarithmic scaling at the classical string side appeared  also  in \ci{sat}.
}
and
the   ``fast long string''  limit
smoothly
interpolating between
 them:
 \be \la{ouo}
   E_0(x \gg 1) =S +  {\sql  \ov \pi}  \ \ln {S\ov J} + { \pi J^2 \ov 2 \sql  \ln {S\ov J} }  + ...
    \ , \ee
 \be \la{ha}
   E_0(x \ll 1) = S+ J + { \l \ov 2 \pi^2 J}
    \ln^2 { S\ov J} -\frac{\lambda^2}{8 \pi^4 J^3}\ln^4{ S\ov J}
    +\frac{\lambda^3}{16 \pi^6 J^5}\ln^6{ S\ov J}
   + ...   \ .   \ee
In this paper we   will  extend the computation \ci{ft1}  of string 1-loop
correction  $E_1= E_1 ( {S\ov \sql}, {J\ov \sql})$
to the energy
of  folded spinning string
to the case of $J\not=0$ in the parameter space region \rf{pp}
 and thus find a closed expression  for
the 1-loop string counterpart of the classical expression \rf{as}.
The result can be written as
 \be
\label{oop}
E_1= {J \ov \sql} \sqrt{1 + x^2} \
 F(x)  +  O(\k^0)   \ , \  \  \   \ \ \ \  \ \ \ \ \k \equiv  {J \ov \sql} \sqrt{1 + x^2} \gg 1 \ ,
\ee
\bea
 &&F(x) =   { 1 \ov   1 + x^2 }
 \bigg[ \, x \sqrt{ 1 + x^2}  -  x^2
 + 2 ( 1 + x^2 ) \ln (1 + x^2)
 \no\\ && ~~~~~~~~~~~~~~~~~~~~~~~~~~~ -\ ( 1 + 2 x^2) \ln [ \sqrt{ 1 + 2 x^2}
    (x + \sqrt{1 + x^2})] \bigg]  \  ,\la{fa}
 \eea
 with $x$ defined in \rf{pp}.
 The counterparts of the ``slow'' and ``fast'' expansions \rf{ouo} and \rf{ha}
 are now
 \begin{equation} \la{ss}
E_1(x \gg 1)=-\frac{3 \ln 2}{\pi} \ln {S \ov J}      + \frac{2 \pi^2
J^2}{\lambda}\frac{\ln \ln {S \ov J}}{\ln {S \ov J}}- \frac{\pi^3
J^4}{\lambda^2}\frac{\ln \ln {S \ov J}}{\ln^3 {S \ov J}}+...\ ,
\end{equation}
\begin{equation}\la{ff}
E_1(x \ll 1) =-\frac{4\l}{3 \pi^3 J^2}\ln^3
\frac{S}{J}+\frac{4\lambda^2}{5 \pi^5 J^4}\ln^5
\frac{S}{J}+\frac{\lambda^{5/2} }{3 \pi^6 J^5}\ln^6
\frac{S}{J}+...  \ .
\end{equation}
\begin{figure}[ht]
\centerline{\includegraphics[scale=0.8]{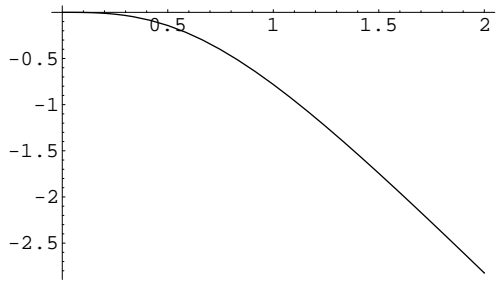}}\caption{Plot of the function
$\sqrt{1+x^2}F(x)$ }
\end{figure}
Figure 1 shows the plot of the 1-loop function
$\sqrt{1+x^2} F(x)$: it  illustrates
that this function  smoothly  interpolates between the   linear $-x$ one
at large $x$  and
$-x^3$ one at small $x$.

Eq.\rf{ss}   provides  further  support for
the universality of the $\ln S$ coefficient  in the quantum string
energy,
 i.e. in the strong-coupling expansion: we reproduce
 the same $ -\frac{3 \ln 2}{\pi} \ln S$ term  as was  found earlier at $J=0$
as a regular  limit of the large $J$ expression.

Checking this universality was one of  our motivations
 in view of   recent
remarkable  results  \ci{bern,bes,kleb,lip} (see also \ci{bhl,oth}).
This universality  of the $\ln S$ coefficient
at $S \gg J \gg 1$  for lowest dimension states in the  $SL(2)$ sector
was argued for at  weak coupling   in  \ci{bgk}  (at one loop) and also more
generally in \ci{es}.
Following \ci{bgk}  and \ci{es}, the work of
 \ci{bes}  was based
 on the  assumption that the coefficient of the
 $\ln S$ term is the same  when computed at
small $J$  and at
large $J$ -- large enough  to be able to apply the asymptotic Bethe ansatz
approach.

The importance of eqs.\rf{as},\rf{fa}  is that  they  allow us to connect the
$\ln S$ terms  to
$J^{-k} \ln^{n} { S \ov J}$
terms
which are ``visible'' in the thermodynamic limit in the weakly coupled gauge
theory spin chain. Indeed, like the ${\l \ov J}\ln^2 { S \ov J}$ term in \rf{ha}
which is reproduced  \ci{bfst} in the 1-loop $SL(2)$ spin chain,
the leading ${\l \ov J^2}\ln^3 { S \ov J}$ term in \rf{ff} agrees precisely
with the leading correction to the thermodynamic limit
of 1-loop gauge theory chain  found in \ci{bgk}  (see  below).
Thus the coefficient of the $\ln S$ term is encoded  as  a limit of the full
 string  1-loop correction determined  from  the string Bethe ansatz \ci{afs,bs,bhl}
  by the BDS \ci{bds} part as well as by the 1-loop  phase of \ci{bt,hl}.\foot{The  results of \ci{kleb,lip} that
   confirm
  the agreement between
 the strong-coupling expansion of the
  integral equation  of \ci{es,bes}
  and the 1-loop result $- {3 \ln 2 \ov \pi } \ln S $    is  a check  of the strong-to-weak coupling
  continuation of the S-matrix phase in \ci{bes}: the starting
 point of  \ci{bes} was   the strong-coupling expansion  of the  phase \ci{bhl}
 which  already encodes this $\ln S$  result.}

To connect  \rf{ha},\rf{ff}  to weakly coupled  gauge theory
let us recall  that
as was emphasised in \ci{bgk},
  in  the   1-loop gauge theory spin chain
 the logarithmic scaling $\l \ln S$  appears universally in the
 thermodynamic limit $J \to \infty$ provided
 $ \ln { S \ov J} \gg J$: it
 takes over  the  leading-order
 ``fast string''
  scaling       $   { \l \ov J } \ln^2 { S \ov J}$
  as one increases the value of the ``gauge-theory''  parameter \ci{bgk}\foot{At
   the transition point
 $ \xi \sim 1$  the semiclassical expansion based on the thermodynamic limit of
 the Bethe equations breaks down \ci{bgk}
 due to the collision of the two cuts at the origin \ci{bfst}.}
 \be \la{ko}
  \xi \equiv  {\pi  \ov \sqrt \l} x   = {  \ln { S \ov J}  \ov  J}   \ . \ee
Motivated by the
analysis of the 1-loop gauge spin chain
 and  also by the strong-coupling string result
\rf{as},
  ref. \ci{bgk} made the following
proposal
 for the all-order behaviour of the  minimal gauge-theory  anomalous dimensions
 in the region $S \gg J \gg 1$
  with fixed $\xi$:
 \be \la{they}
 E= S+J+  J \sum_{n=1}^\infty  c_n(\x)  (\l \x^2)^n \  , \ee
 where
 \be \la{the}
 c_n(\x \ll 1 ) = c_{n0} + c_{n1} \x + c_{n2} \x^2  + ...\ ,
 \ee
 \be  \la{thet}
  c_n(\x \gg  1 ) =  {a_n \ov \x^{2n-1}} + ... \ .
 \  \ee
 The terms in the ``fast string''  \ci{ft1,ft2}  or
  BMN-type scaling \rf{the} with $\x \ll  1$
 which multiply  $c_{n1}$, $ c_{n2}$,...  scale  as ${1 \ov J}, {1\ov J^2}, ...$
and thus
represent corrections to the thermodynamic $J\to \infty $ limit.
The   large $\x $ scaling  behaviour of $c_n$  in \rf{thet}
   translates into the familiar perturbative $\ln S$ scaling
   \be
   E =S + J +    \  (a_1 \l + a_2 \l^2 + ...)\ln S + ...  \ .  \ee
 The coefficients in \rf{the} were assumed in \ci{bgk}
  to have no  additional
 dependence on $\l$.
 
However,  it is known
   \ci{bt} that   in  the  {\it  strong coupling}  limit as described by string theory, 
   where $E$ is expressed in terms of $\l \gg 1$  and the semiclassical string parameter 
   $x$ in \rf{ko} (in which one can subsequently expand assuming $x  < 1$) 
   the  ``fast  string''  scaling  (i.e.  \rf{they} written  in terms of $x$) 
 breaks down starting with  $x^6$ (or ``3-loop'')  term. 
 As we shall explain below, eq.\rf{ff} provides
 a direct indication  of that, 
in complete  analogy with the results of \ci{bt,sz,hl}  for circular  string solutions.
 The strong-coupling corrections to the dressing phase  \ci{hl} translate \ci{bhl,bes} 
 into the weak-coupling correction which should 
 contribute starting with  4-loop term in the 
 weak-coupling expansion in \rf{they}.
 
This suggests   that   \rf{they}  viewed as an exact expression for the energy 
interpolating between weak and strong coupling regions 
should  be modified  following
\ci{bt,mtt2,bes}:
most of the coefficients $c_n$ in \rf{they} except for the few leading ``protected'' ones 
 should develop additional dependence on
$\l$, allowing, in particular, for the ``3-loop'' coefficient 
 $c_{30}$ to have different limiting
values at $\l\to 0$ and $\l\to \infty$, 
Explicitly, one may expect (see also below)
\bea
 E= S+J\bigg[1 &+&  { \l \ov J^2} \ln^2 { S  \ov J}\
  ( c_{10} + c_{11} { \ln { S\ov J} \ov J}   + c_{12} { \ln^2 { S\ov J} \ov J^2} +  ...)
      \no   \\
  &+&  { \l^2 \ov J^4} \ln^4 { S  \ov J}\
  ( c_{20} + c_{21} { \ln { S\ov J} \ov J}   + c_{22}(\l)  { \ln^2 { S\ov J} \ov J^2} +  ...)
 \la{tey}  \\
  &+&  { \l^3 \ov J^6} \ln^6 { S  \ov J}\
  \big( c_{30}(\l) + c_{31}(\l)  { \ln { S\ov J} \ov J}   + c_{32}(\l)
   { \ln^2 { S\ov J} \ov J^2} +  ...\big)
   + ... \bigg] \ , \no
   \eea
The  functions $c_{nk}(\l)$  should have regular expansion at  weak coupling, e.g., \foot{
Note that there is an ambiguity in how one splits  corrections between different terms, 
e.g., $\l^2 c_{22}(\l)$ and $\l^3 c_{30}(\l)$ which both multiply  
$ J^{-6} \ln^6 { S  \ov J}$  in  the brackets in \rf{tey}.
}
\be \la{ccc}
c_{30}(\l\ll 1 ) =c_{30,0} + c_{30,1} \l +  O(\l^2)  \ , \ \ \ \ {\rm etc.}  \ \ee
The weak-coupling coefficient  $c_{30,0}$ should be different
from  the strong-coupling  one  ${ 1 \ov 16 \pi^2}$ in \rf{ha}
(manifesting ``3-loop disagreement'' \ci{call,ss})
while the first violation of the ``semiclasical'' scaling  $c_{30,1}\not=0$
at weak coupling should thus appear at {\it four}  loops.
 The weak-coupling
expansion of the dressing phase proposed in  \ci{bes}  suggests  that 
a direct gauge-theory computation
of this coefficient should give
  $c_{30,1}\sim \zeta(3)$.

Let us now compare \rf{they},\rf{tey} with the string theory predictions $E_0 +E_1$
 in \rf{ouo},\rf{ha} and    \rf{ss},\rf{ff}
for the leading strong-coupling corrections.
The leading term in \rf{ff} has the same structure as the $c_{11}$ term in \rf{tey}.
Remarkably, like $c_{10}={1 \ov 2\pi^2}$
that matches \ci{bfst} the coefficient of the leading $J^{-1} \ln^2 {S \ov J}$ correction in
\rf{ha},  its coefficient in \rf{ff} is exactly the same
as the
1-loop gauge theory  coefficient
\be c_{11}=  -{4 \ov 3 \pi^3}   \ee
  computed
as the leading  correction to  the thermodynamic
limit in the $SL(2)$ Heisenberg chain
 in \ci{bgk}.\foot{The result
for $c_{11}$ is given at the
end of sect. 3.2 in \ci{bgk} (here $s={1\ov 2}$ since the string states we consider
are dual to  scalar operators).
 We thank A. Belitsky for pointing this out
  to us.}

Looking at higher orders,
 the $c_{20}$ term in \rf{tey} has the same form and should also have the same coefficient
 as the $J^{-3} \ln^4 { S\ov J}$ term in the classical string energy
 \rf{ha}, i.e. $c_{20}= - { 1 \ov 8 \pi^4}$.
We also observe that
  the  absence of the $J^{-3} \ln^4 { S\ov J}$ term in
  the string 1-loop correction  \rf{ff} is in full agreement with
 \rf{tey}: the subleading 2-loop  $c_{21}$ term there scales as $ J^{-4} \ln^5 { S\ov J}$
 and  should
  have the same coefficient as the corresponding term in \rf{ff},
  i.e.  $c_{21}= {4 \ov 5 \pi^5}$.\foot{The
   coefficients $c_{20}$ and $c_{21}$
  were not yet computed on the gauge theory side. It would be nice to
  check the above values directly  by  extending the methods of \ci{bgk}
  to 2-loop order.}

 The non-renormalization of the few leading 1- and 2-loop coefficients in \rf{tey}
 in going from weak to strong coupling is of course
 expected on the basis of  consistency  with previous results
 for similar circular string solutions.
  The 1-loop  and 2-loop  leading and  the first subleading \ci{btz,hla,szz,mtt1,gk,mtt2}
  corrections  to the thermodynamic limit
 should match  precisely  between  gauge and string theory: the effect of the 
  non-trivial
   phase in the Bethe ansatz should first become visible in leading thermodynamic limit
   and strong-coupling expansion  only
  starting
  with  terms of $\l^3$  order in ``semiclassical''expansion like \rf{tey}.\foot{The presence of higher-order
  corrections
in the strong-coupling expansion of the dressing phase \ci{bhl}
corresponding to 2- and higher loop quantum string corrections and
translating  into the presence of further subleading terms in the strong-coupling expansion  of the coefficient $c_{30}$ in \rf{tey} (see \rf{hi} below)
appears to imply that, in contrast to $c_{20}$ and $c_{21}$,
  the ``$1/J^2$''  2-loop coefficient $c_{22}$
should not be protected when going from weak to strong coupling.
Same should be true  also for the  ``$1/J^4$''  1-loop coefficient $c_{14}$
in \rf{tey}, etc. Similar remark should apply to the
expansion of the energy of other semiclassical solutions (note that in the near-BMN case \ci{call} the order
at which weak-strong coupling non-renormalization  should fail is shifted by one power  of $1/J$, 
see sect. 6 in  \ci{mtt2}).
It  would be important to  study  in detail how the proposal for the dressing phase 
in \ci{bhl,bes} extending \ci{bt,hl} to all orders 
 modifies the ``semiclassical'' expansion  originally  conjectured in \ci{ft2,ft3}.
 We are grateful to N. Beisert for a discussion of this  issue.}



Indeed, combining the $J^{-5} \ln^6 { S\ov J}$ terms in \rf{ha} and \rf{ff} and comparing them to \rf{tey} we conclude that
as in \ci{bt} the presence of ``non-analytic'' $\l^{5/2}$ term in string 1-loop correction \ci{bt,sz} implies the
renormalization of the ``3-loop'' coefficient $c_{30}$ in \rf{tey}:
 \be \la{hi}
  c_{30}(\l\gg 1 ) = 1 + { 16 \ov 3 \sql} +  O({ 1 \ov \l}) \ .
 \ee
As was found in \ci{bt}
 on the examples of circular solutions in $SL(2)$ \ci{ptt} and $SU(2)$ \ci{fpt} sectors\foot{Same result is found  also for the circular
 solution  \ci{ft3} in  the $SU(3)$  sector \ci{kf}.}
 the non-analytic $\l^{5/2}$ correction in the string 1-loop energy
 should universally  account for the  difference
 between  the string and gauge   predictions for the coefficient of the $\l^3$ term,
 corresponding  to the function
 \be \la{pha}
 {\rm c}_2(\l)=1 - { 16 \ov 3 \sql} +  O({ 1 \ov \l})\ee
  in
  the dressing phase in the Bethe ansatz which should interpolate between
  1  \ci{afs} at strong coupling  and  0 \ci{bds} at weak coupling.
  For example, in the case of $J_1=J_2$ circular solution in the $SU(2)$ sector
  one finds from the classical energy $E_0= \sqrt{J^2 + \l m^2}$
  and the 1-loop correction  in  \ci{bt} (see also appendix C in \ci{mtt2})
   that the string prediction is
   $E_0+E_1 = ... + {\l^3 m^6 \ov J^5} (1 - { 16 \ov 3 \sql}) + ... $.
  This  is consistent with the above expression for the
  phase function
  \rf{pha} since it is known  that in this case the BDS ansatz
  gives zero contribution at order $\l^3\ov J^5 $ \ci{ss}
  (for the discussion of the case of circular solution with $J_1 \not= J_2$ see
  appendix A in \ci{mtt2}).
  Similarly, our present string result \rf{hi} will be consistent with
  the universal  phase \rf{pha}  provided the weak-coupling gauge-theory coefficient
  $c_{30,0}$ in \rf{ccc} is twice its strong-coupling limit, i.e. the  classical string value in \rf{ha}, i.e.
  \be \la{pre}
  c_{30,0} = { 1 \ov 8 \pi^6}   \ . \ee
  Again, it would be interesting to  confirm this prediction by a direct computation on the gauge theory side, thus checking again
  the  universal origin of  the 1-loop phase \ci{bt,hl}
 that corrects the string Bethe ansatz S-matrix \ci{afs}.

We conclude that the structure of  our 1-loop string result \rf{oop}
appears to be  perfectly consistent
with all so far  known facts about the strong  coupling expansion of the
 Bethe ansatz.


\renewcommand{\theequation}{2.\arabic{equation}}
 \setcounter{equation}{0}

\section{1-loop correction to folded $(S,J)$ string energy in
 the ``long string'' approximation}

Let us now describe the details of the computation leading to \rf{oop}.
What follows will be  heavily  based on the results of \ci{ft1}.

Let us first recall the form of the folded string solution found  in \cite{ft1}
(which generalizes the one in \ci{dev,gkp})
\begin{equation}
t=\kappa \tau,\quad \rho=\rho(\sigma), \quad \phi=\omega \tau, \quad \varphi=\nu \tau
 \ .
\end{equation}
Here $t$ is global time, $\r$ is the radial coordinate in $AdS_5$ and  $\phi$
 and  $\varphi$ are angles in  $AdS_5$ and $S^5$ respectively.  $\rho$ is determined from
\begin{equation}\la{rhop}
\rho''=(\kappa^2-\omega^2) \sinh \rho \cosh \rho\ , \ \ \ \ \ \  \ \ \
\rho'^2=\kappa^2 \cosh^2 \rho -\omega^2 \sinh^2 \rho -\nu^2 \ .
\end{equation}
The periodicity condition on $\rho(\sigma)=\rho(\sigma+2 \pi)$ is satisfied by
considering a folded string configuration. For $0\leq \sigma <
\pi/2$ the function $\rho'$ increases from $0$ to its maximal
value $\rho_0$, while for $\pi/2< \sigma <\pi
$, $\rho$ decreases from $\rho_0$ to $0$.
It is useful to introduce the parameter $\eta$ as
\begin{equation}
\coth^2 \rho_0=\frac{\omega^2-\nu^2}{\kappa^2-\nu^2}=1+\eta, \quad\quad
\quad \eta>0
\end{equation}
The periodicity condition implies
\begin{equation}
\sqrt{\kappa^2-\nu^2}=\frac{1}{\sqrt{\eta}}
  \ {}_2F_1\bigg(\frac{1}{2},\frac{1}{2};1;-\frac{1}{\eta}\bigg)
\end{equation}
The non-zero  charges are the energy $E$ and two angular
momenta $S$ and $J$
\begin{equation}
E= \sqrt{\lambda}\kappa \int_0^{2 \pi} \frac{d \sigma}{2 \pi}
\cosh^2 \rho= \sqrt{\lambda}\mathcal{E}, \quad S=
\sqrt{\lambda}\omega \int_0^{2 \pi} \frac{d \sigma}{2 \pi} \sinh^2
\rho=\sqrt{\lambda}\mathcal{S}, \quad J=\sqrt{\lambda}\nu
\no
\end{equation}
where $
\mathcal{E}=\kappa+\frac{\kappa}{\omega}\mathcal{S}$
and
\begin{equation}
\mathcal{E}=\frac{\kappa}{\sqrt{\kappa^2-\nu^2}}\frac{1}
{\sqrt{\eta}}
 \ {}_2F_1\bigg(-\frac{1}{2},\frac{1}{2};1;-\frac{1}{\eta}\bigg), \quad
\mathcal{S}=\frac{\omega}{\sqrt{\kappa^2-\nu^2}}\frac{1}{2\eta
\sqrt{\eta}} \
{}_2F_1\bigg(\frac{1}{2},\frac{3}{2};2;-\frac{1}{\eta}\bigg)
\end{equation}
The above hypergeometric  functions can be expressed in terms of standard elliptic
integrals (see \ci{bfst}).

Below we will be  interested in the particular
 ``long string'' limit
$\rho_0\rightarrow \infty$, i.e. $\eta \ll 1$, in which
the computation  can be drastically simplified \ci{ft1},
 assuming one is interested in the leading large $\k$ correction to the
 string energy.
 In this limit
\begin{equation}\la{longstring}
\kappa^2 \approx \nu^2+ \frac{1}{\pi^2}\ln^2 {\eta}, \quad\quad
\omega^2 \approx \nu^2+ \frac{1}{\pi^2}(1+\eta)\ln^2
{\eta}, \quad\quad \mathcal{S}\approx - \frac{2 \omega}{\eta \ln
\eta}\ ,
\end{equation}
i.e. $\kappa\approx \omega$ and
$\mathcal{S}\gg 1$.  To describe both the ``slow long string'' and the ``fast long string'' limits  we consider a special scaling of $\nu={J\ov \sql}$) such that
\be \la{scaling}
 u\equiv  { \nu \ov \k}  \ee
is kept fixed in the long string limit $\eta\to 0$. By using  (\ref{longstring}) we see that this scaling implies that
\bea
\kappa \approx -{\ln \eta \ov\pi \sqrt{1-u^2}}\,,\quad \quad
\mathcal{S}\approx - \frac{2 \kappa}{\eta \ln
\eta} \approx  \frac{2}{\pi \sqrt{1-u^2}}{1\ov\eta }\,,
\eea
and, therefore, the parameter $\k$ is expressed through the spin $\S$ and the parameter $u$ as follows
\bea\la{kap}
\kappa \approx {\ln ( {\pi\ov 2}\mathcal{S} )    +{1\ov 2} \ln(1-u^2) \ov\pi \sqrt{1-u^2}}\,.
\eea
Taking into account that  $u={\nu\ov \k}$, we find that $u$ as a function of $\nu$ and $\S$ is given by the solution of the following equation
\bea\la{u}
u \approx {\pi\nu\sqrt{1-u^2}\ov \ln ({\pi\ov 2}\mathcal{S}) +{1\ov 2} \ln(1-u^2)}\,.
\eea
In the large $\mathcal{S}$ limit, and $u<1$, and not approaching 1, the solution of this equation takes the form
\bea \la{uj}
u \approx {\pi\nu\ov \ln \S }{1\ov \sqrt{1+({\pi\nu\ov \ln \S})^2}}\,,
\eea
where we replaced $\ln ( {\pi\ov 2}\mathcal{S}) $ by $\ln \S$.
It is worth noting  that we do not assume here that $u$ is small. This formula is valid for large $\mathcal{S}$, and
any finite $u<1$, e.g.,  for $u=1/2$. Since $u$ is kept fixed in the large $\S$ limit, eq.\rf{uj}  also implies that the ratio $\nu/ \ln \S  \sim J/\ln S$ is fixed too. Taking into account that $\k = {\nu\ov u}$, and  that in the long string limit $E- S \approx  \sqrt\l \kappa$, we get the following expression for the classical energy
\bea\la{clen1}
E_0 - S \approx  {\sqrt\lambda\ov\pi} \ln \S \sqrt{1+({\pi\nu\ov \ln \S })^2} = \sqrt{J^2 + {\l\ov\pi^2} {\ln^2{S\ov\sqrt\l}}}\,\ .
\eea
Expanding this  in powers of ${\pi\nu\ov \ln \S }$, we recover the expression  derived in \cite{ft1}.

To consider the ``fast long string'' case with $\S\gg \nu\gg \ln {\S\ov\nu}$ corresponding to the limit  $u\to 1$, we introduce the variable $y$ such that
\bea
 1- u^2 = {y^2\ov\pi^2\nu^2}\
\eea
and take the large $\nu$ limit assuming $y\ll  \nu$. Then the equation (\ref{u}) for $u$ takes the form
\bea
\sqrt{1-{y^2\ov\pi^2\nu^2}}\approx {y \ov\ln ({\mathcal{S}\ov\nu} y )}\,,
\eea
and for $S\gg \nu$ we get
\bea
y \approx {\ln {\mathcal{S}\ov\nu}\ov \sqrt{1+{\ln^2{\S\ov\nu}\ov\pi^2\nu^2}} }\,.
\eea
This leads to the following expression   for the energy of the fast long string
\bea\la{clen2}
E_0-S \approx  \sqrt\l \nu\sqrt{1+{\ln^2{\S\ov\nu}\ov\pi^2\nu^2}}= \sqrt{J^2 + {\l\ov\pi^2} {\ln^2{S\ov J}}}
\eea
first derived in \cite{bgk}  and already quoted above in \rf{as}.

\bigskip

To compute the $1$-loop string correction to the
classical energy we  shall start with  the bosonic fluctuation
action in conformal gauge as found in  \cite{ft1}: $ I=I_1+I_2$, where
\begin{eqnarray}
I_1&=&-\frac{1}{4 \pi}\int d^2 \s  \bigg[-\partial_{a} \bar{t}
\partial^{a} \bar{t}-\mu_t^2 \bar{t}^2+\partial_{a} \bar{\phi}
\partial^{a} \bar{\phi}+ \mu_\phi^2 \bar{\phi}^2+ \partial_{a} \bar{\rho}
\partial^{a} \bar{\rho}+ \mu_\rho^2 \bar{\rho}^2 \  \nonumber \\
&+&  4 \bar{\rho}(\kappa \sinh \rho\
\partial_{0}\bar{t}-\omega \cosh \rho\ \partial_{0}
\bar{\phi})\bigg]\ ,  \label{ads1}
\end{eqnarray}
\begin{equation}\la{ad}
I_2=-\frac{1}{4 \pi} \int d^2 \s  \bigg[\partial_{a} {\beta}_i
\partial^{a} {\beta}_i+ m_\beta^2 \beta_i^2+\partial_{a} \bar{\varphi}
\partial^{a} \bar{\varphi}+\partial_{a} \psi_s
\partial^{a} {\psi}_s+ \nu^2 {\psi}_s^2\bigg] \ ,
\end{equation}
and
$$
\mu_t^2= 2 \rho'^2-\kappa^2+\nu^2, \quad \mu^2_\phi=2
\rho'^2-\omega^2+\nu^2, \quad \mu^2_\rho= 2
\rho'^2-\kappa^2-\omega^2+2\nu^2, \quad m^2_\beta= 2 \rho'^2+\nu^2
$$
Here $\beta_i$ ($i=1,2$) are fluctuations of two angles of $AdS_5$
transverse to $AdS_3$ in which the string is moving; $\psi_s$
($s=1,2,3,4$) are fluctuations of four $S^5$ directions transverse
to the center of mass motion.

Solving for the spectrum of
fluctuations in general is a difficult task since $\rho$  is a
non-trivial (elliptic) function of $\sigma$.  Fortunately, in the long string limit
 the fluctuation action can be brought
to more tractable  form.
 To this end  we may first perform a field redefinition
\begin{equation}
\chi=\bar{\phi} \cosh \rho- {\kappa\ov\omega}\bar{t} \sinh \rho, \quad\quad\quad \zeta=-
\bar{\phi} \sinh \rho+  {\kappa\ov\omega}\bar{t}\cosh \rho\,
\end{equation}
which simplifies the cross-term on the second line in (\ref{ads1})
when $\k \approx \omega$ as is true  in the long-string limit
\rf{longstring}  when  $\eta\to 0$ with $u={\nu\ov\kappa}$ fixed.
 It was shown in \ci{ft1} (see section 6.2 there) that in this limit
$\rho'$ is approximately constant and is equal to
\begin{equation}\la{apr}
\rho'  \approx \pm \sqrt{\kappa^2-\nu^2}\ ,\ \ \ \ \ \
\ \ \ \ \ \ \ \rho'' \approx  0 \ .
\end{equation}
 except at the turning points
$\sigma=\frac{\pi}{2},\frac{3 \pi}{2}$ where $\rho'=0$.
As was argued in \ci{ft1} (and as in a similar situation in  \cite{mtt3}),
contribution of these isolated points
 may  be ignored in the computation
of the spectrum to leading order in large $\k$.
Then the action (\ref{ads1}) expressed in terms of the rotated coordinates $\chi$ and $\zeta$ has constant coefficients
\begin{equation}
I_1\approx -\frac{1}{4 \pi} \int d^2 \s
\bigg[-\dot{\chi}^2+\chi'^2+\dot{\zeta}^2- \zeta'^2+ 4
\sqrt{\kappa^2-\nu^2}\chi' \zeta- 4 \kappa
\dot{\chi}\bar{\rho}-\dot{\bar{\rho}}^2+ \bar{\rho}'^2\bigg] \
\end{equation}
and thus the spectrum of characteristic  frequencies is   readily
computable. We find that one combination of the $AdS_3$ modes
$\chi, \zeta,\bar \rho$  is massless, and thus,
like  the massless mode  $\bar \varphi$ in \rf{ad}, it
does not produce  nontrivial contribution to string energy.\foot{The
contribution of these two massless decoupled modes is cancelled against
 the conformal gauge ghost
contribution \ci{ft1}.}
The remaining two modes
have the following frequencies
\begin{equation}
\Omega_{\pm n}=\sqrt{n^2+2\kappa^2 \pm 2 \sqrt{\kappa^4+n^2 \nu^2}} \ , \ \ \ \ \ \
n=0, \pm 1, \pm 2 , ... \ .
\end{equation}
In addition to these two $AdS_3$ modes
there are also  $2$ transverse $AdS_5$ bosonic modes  with
mass $m^2_\beta\approx 2 \kappa^2-\nu^2$    and 4 transverse $S^5$  bosonic
frequencies with mass $\nu^2$.

As was shown  in \ci{ft1}, after  the $\kappa$-symmetry gauge fixing
(and before making any approximations)
the quadratic fermionic part of the \adss superstring action reduces to the
standard  action
for $4+4$ $2d$ Majorana fermions
with $\rho(\s)$-dependent masses $m_F= \pm  \sqrt{ \rho'^2 + \nu^2}$.
 In the ``long string'' approximation the square of these masses
 becomes approximately constant  and equal to  $m^2_F\approx \k^2$, so that
 the fermionic frequencies are
\begin{equation}
\Omega_{F n}\approx  \sqrt{n^2+\kappa^2} \ .
\end{equation}
Therefore,  the $1$-loop correction to the string energy
can be written as\foot{The zero-mode contribution $K_0$
in a similar equation (6,6) in \ci{ft1} was omitted since it gave only
subleading contribution
to the energy. Same will apply here.}
\begin{eqnarray}
\label{sum1} E_1\approx
\frac{1}{2\kappa}\sum_{n=-\infty }^{\infty}   K_n =
\frac{1}{\kappa}
   \bigg(\sum_{n=1}^{\infty}K_n + {1 \ov 2}  K_0 \bigg) \ ,
\end{eqnarray}
where
\begin{equation}
K_n= \Omega_{+n}+\Omega_{-n} +   2 \sqrt{n^2+2 \kappa^2-\nu^2}+4
\sqrt{n^2+\nu^2} -8 \sqrt{n^2+\kappa^2} \ .
\end{equation}
It is easy to  check that this
sum is convergent in the UV.
For $\nu=0$  this sum  reduces to the expression  of \ci{ft1} found in the
$J=0$ case in the static gauge. This confirms  that the computations in
the static gauge  and the conformal gauge agree as they
should.

Here  we are interested in the  value of the sum \rf{sum1} in the scaling limit (\ref{scaling}) when
\be \la{kaa}
\kappa = {\nu\ov u}\  \gg 1 \ , \ \ \ \ \ \ \ \
 u= {\rm fixed} \ . \ee
As in \ci{ft1} (see also \ci{mtt2,sak})
the leading
large $\kappa$  asymptotics of the sum in \rf{sum1}
 can be found by replacing it by an integral
\begin{eqnarray}
E_1&\approx &\kappa \int_0^{\infty}dp\ \bigg[\sqrt{p^2+2+2 \sqrt{1+u^2 p^2}}
+ \sqrt{p^2+2-2 \sqrt{1+u^2 p^2}}
\no \\
 &+&
2\sqrt{p^2+2-u^2}+4 \sqrt{p^2+u^2}
-8\sqrt{p^2+1}\ \bigg]\ +O(\kappa^{0}) \ . \label{sum2}
\end{eqnarray}
This integral   happens  to be essentially the same
  as
in the case of the 1-loop correction to the energy of the 
circular  $J_1=J_2$ string solution in $SU(2)$ sector \ci{ft2,art}
considered in \ci{bt} and  in Appendix C of \ci{mtt2}.
The precise relation between  the parameters is as follows:
$\kappa$ in \ci{mtt2} is $\nu$ here, and  the winding number $m$  in \ci{mtt2}
is replaced by the imaginary value $i \sqrt{ \k^2 - \nu^2}$ here
(recall that the circular solution of \ci{ft2} is unstable
unless $m^2$ is formally less than 1 or negative; the 
 folded string solution considered here is stable).  Finally, the rescaled angular momentum 
 ${\cal J} = \sqrt{ \kappa^2 - m^2}$ in \ci{mtt2} corresponds   then $\kappa$ here.

The reason  why these  two 1-loop expressions 
are related in this curious way   is as follows.
In the long string limit of the folded string solution 
eq. \rf{apr} implies $\rho \approx \mm \sigma, \ \mm \equiv \pm  \sqrt{ \k^2 - \nu^2}$, 
i.e. linear in $\sigma$. But then the  above $AdS_3\times S^1 $ 
($ds^2 = d\r^2 - \cosh^2 \r dt^2 + \sinh^2 \r d \phi^2 + d \vp^2 $)  
configuration  with $\omega \approx \k$, i.e.
    $t= \k \tau, \ \p \approx \k \tau, \ \r\approx \mm \sigma, \ \vp =\nu \tau$ 
is related by a formal analytic continuation as in \ci{bfst} to 
the $J_1=J_2$ circular string solution  in $R \times S^3$ 
($ds^2 =- dt'^2 +  d\theta^2 + \cos^2 \theta d\vp'^2_1 + \sin^2 \theta d \vp'^2_2$)   
 taken  in its original (unrotated, cf.  \ci{art}) form given in \ci{ft2}:
  $t'= \k' \tau, \ \theta = m' \s,\   \vp_1= \vp_2 = w'  \tau, \ 
    w'= \sqrt{ \k'^2 - m'^2}$.\foot{Under the continuation $t \to \  \vp'_1, \ 
    \rho \to i \theta, \ \p \to \vp'_2, \ \vp \to t'$ and one is to change the overall sign of the 
    action.}

The  evaluation of the integral \rf{sum2} 
is thus done in  the same way as in \ci{mtt2} 
-- by introducing an UV cutoff, doing individual
integrals and then taking the cutoff to infinity.
Using the identity
$$
\sqrt{p^2+2+2 \sqrt{1+u^2 p^2}} +\sqrt{p^2+2-2 \sqrt{1+u^2
p^2}}=\sqrt{4 u^2+ (p+\sqrt{p^2+4-4 u^2})^2}
$$
and  changing the  variable in the corresponding integral ($p \to z=p+\sqrt{p^2+4-4 u^2}$)
we end up with
\begin{eqnarray}
&& E_1\approx  -  {\nu\ov u}\bigg[ 1-u^2 - \sqrt{1-u^2}
 \no \\&&~~~~~~~~~~~~~~+(2
- u^2) \ln [ \sqrt{ 2- u^2} (1+ \sqrt{1-u^2})]  +  2
u^2 \ln u\bigg] +  O(\k^0) \ . \label{1loop}
\end{eqnarray}
Written in terms of $x$ in \rf{pp} and $\nu= {J\ov  \sql }$
 ($\k= {\nu \ov u}, \  u = {1 \ov \sqrt{1 + x^2}}$)
  this is the same
expression as  was given earlier   in \rf{oop},\rf{fa}.

\renewcommand{\theequation}{3.\arabic{equation}}
 \setcounter{equation}{0}

\section{  Concluding remarks              }

In this paper
we computed the one-loop string correction to the energy  of folded $(S,J)$ string
in a
special ``long string''
  limit, confirming the universality of the $\ln S$ coefficient.

   It would be interesting  to reproduce this  one-loop correction by
   starting with  the quantum string Bethe ansatz.  It would give an additional
    nontrivial check of the one-loop correction \cite{hl} to the AFS phase \cite{afs}.
     The new scaling in  $\nu = { J \ov \sql}$ in
   the large $\S= { S \ov \sql} $ limit
   (preserving  the leading
$\ln S$ behavior of the string energy but leading  to a
nontrivial dependence on the parameter $u$ in \rf{u})
which we considered  in section 2
may   serve  as a regulator of the complicated singular integral equations describing
 the two-spin folded string case in the BA approach.


As we mentioned in section 1, there are few perturbative gauge-theory
computations of the leading coefficients in \rf{tey} that  remain to be
done in order to check explicitly the correspondence  between
our results and the expected  ``dressed''  Bethe ansatz picture.
 It would be interesting  also to understand the meaning of the $\ln \ln S$
 terms appearing in the subleading terms in strong-coupling result \rf{ss}
 on  the gauge theory side.

On the string-theory  side, it would be very important to
find the  2-loop string-theory counterpart of the 1-loop correction discussed here.
This computation may be feasible in the ``long-string'' limit where
the string fluctuation Lagrangian is expected to simplify. This would lead  to
many new checks of the proposed dressing phase \ci{bhl,bes}.

Among other possible string-theory generalizations,
let us mention the
computation of 1-loop correction to the energy of long rotating strings
with  $n$ spikes \ci{kru} (with small or large $n >2$)
and also the computation of 1-loop   energy  of strings with two spins $(S_1,S_2)$ in $AdS_5$
 \ci{ptt2}. This may shed further light on the universality of the $\log S$ scaling
 and the properties of the quantum string Bethe ansatz.

\bigskip
\bigskip

\section*{Acknowledgments }

We are  grateful to N. Beisert,  A. Belitsky,  I. Klebanov, R.
Roiban and M. Staudacher for useful  discussions. S.~F.~was
supported in part by the EU-RTN network {\it Constituents,
Fundamental Forces and Symmetries of the Universe}
(MRTN-CT-2004-005104). A.T. thanks the Physics Department of  The Ohio
State University for its support.
   A.A.T.
acknowledges the support of   PPARC, EU-RTN network  MRTN-CT-2004-005104 
 and  INTAS 03-51-6346 grants,    and the RS Wolfson award.

\end{document}